\documentclass[aps,showpacs]{revtex4}
\usepackage{amssymb}
\usepackage[dvips]{graphicx}
\usepackage[english]{babel}
\usepackage{indentfirst}
\usepackage{amsxtra}
\usepackage{amsmath}
\usepackage{supertabular}
\usepackage{multirow}
\usepackage [mathcal]{eucal}
\def\beq{\begin{equation}}
\def\eeq{\end{equation}}
\def\beqn{ \begin{eqnarray} }
\def\eeqn{ \end{eqnarray} }
\def\s1s2{{ \boldsymbol{\sigma}(1) \cdot \boldsymbol{\sigma}(2) }}
\def\t1t2{{ \boldsymbol{\tau}(1) \cdot \boldsymbol{\tau}(2)  }}
\newcommand{\doc}{{\cal D}}
\newcommand{\rbe}{{\cal R}}
\newcommand{\nst}{$N^*$~}
\newcommand{\npi}{$N(\pi)$~}
\newcommand{\nnu}{$N(\nu)$~}
\newcommand{\bsigma}{\mbox{\boldmath $\sigma$}}
\newcommand{\btau}{\mbox{\boldmath $\tau$}}
\newcommand{\half}{\frac{1}{2}}
\newcommand{\what}[1]{\widehat #1}

\newcommand{\threej}[6]{ \left( \begin{array}{ccc}
                               #1 & #2 & #3 \\
                               #4 & #5 & #6 
                             \end{array}
                        \right) }

\begin{document}
\noindent
\title{Evolution of the pygmy dipole resonance in nuclei
with neutron excess}
\author{G. Co', V. De Donno  and C. Maieron}
\affiliation{Dipartimento di Fisica, Universit\`a del Salento and,
 INFN Sezione di Lecce, Via Arnesano, I-73100 Lecce, ITALY}
\author{M. Anguiano and A. M. Lallena}
\affiliation{Departamento de F\'\i sica At\'omica, Molecular y
  Nuclear, Universidad de Granada, E-18071 Granada, SPAIN}

\date{\today}

\begin{abstract}
The electric dipole excitation of various nuclei is calculated with a
Random Phase Approximation phenomenological approach. The evolution of
the strength distribution in various groups of isotopes, oxygen,
calcium, zirconium and tin, is studied.  The neutron excess produces
$E1$ strength in the low energy region. Indexes to measure the
collectivity of the excitation are defined. We studied the behavior of
proton and neutron transition densities to determine the isoscalar
or isovector nature of the excitation. We observed that in
medium-heavy nuclei the low-energy $E1$ excitation has characteristics
rather different that those exhibited by the giant dipole
resonance. This new type of excitation can be identified as pygmy
dipole resonance.
\end{abstract}

\maketitle

\section{Introduction}
\label{sec:intro}

There are experimental evidences that in nuclei with neutron excess,
in addition to the well known Giant Dipole Resonance (GDR), a new type
of dipole resonance appears \cite{adr05,rye02,har04}.  Since this
resonance has smaller strength than that of the GDR and exhausts only
a small fraction of the energy weighted sum rule, it is called Pygmy
Dipole Resonance (PDR).  The PDR appears at lower energy with respect
to the GDR but it is not its low-energy tail, since it has an
isoscalar ($IS$) character and it is dominated by neutron excitations,
while the GDR has isovector ($IV$) character with almost equal
contribution of proton and neutron excitations. 

The existence of a new type of resonance in nuclei is an interesting
subject by itself. In the present case the interest is also related to
the fact that the presence of the PDR may have some relevant consequences
in the stellar r-process production of exotic nuclei \cite{gor98}.
 
We have studied how the PDR emerges when neutrons become more numerous
than protons. The nuclear model adopted in our calculation is the
traditional discrete Random Phase Approximation (RPA). Non
relativistic \cite{cat97,cha94,oro98} and relativistic
\cite{vre01a,vre01b,pie06} RPA approaches have been used in the past to
investigate the PDR. Studies of PDR have been also done with more
elaborated nuclear models containing pairing
\cite{kam98,gor02,gia03,gor04} and spreading widths
\cite{col01,sar04,tso04a,tso04b,tso06,lit07,tse07}.  Usually these
calculations have been done to make detailed investigations of the PDR
characteristics in a limited set of isotopes. Our goal here is to
search for general trends of the PDR in various nuclei belonging to
different regions of the nuclear isotope table, from oxygen to
lead. For this purpose we have defined two indexes which enable us to
distinguish between PDR and GDR.  We tested the validity of these
investigation tools on the $^{208}$Pb nucleus where we found a
resonance with all the features we attribute to the PDR at about 7.7
MeV, against an experimental evidence around 7.35 MeV \cite{rye02}.
We applied our method to isotopic chains of oxygen, calcium, zirconium
and tin. We clearly identify the emergence of the PDR in the isotopes
with neutron excess.

The basic features of our model are presented in Sect. \ref{sec:model}
where we also make a critical discussion of its limits. We present 
our results in Sect. \ref{sec:resu} and in Sect. \ref{sec:conc}
we summarize our work and draw our conclusions. 

\section{The model}
\label{sec:model}
In our work we adopted the phenomenological RPA approach as proposed
and used by the J\"ulich group \cite{spe77,rin78,spe80}.  The
single-particle (s.p.) basis is constructed on a Woods-Saxon well
whose parameters are fixed in order to reproduce at best the
s.p. energies around the Fermi surface and the charge
distributions. We used the parameters of the Woods-Saxon potential
given in Ref. \cite{ari07}. Only the parameters of the $^{90}$Zr and
$^{132}$Sn nuclei are new, and they are presented in
Table \ref{tab:WS}. We used the expression of the Woods-Saxon potential
given in Ref. \cite{ari07}.  In our RPA calculations we used
experimental s.p. energies when available.

\begin{table}[h]
\begin{center}
\begin{tabular}{llccccccc}
\hline
 &   & $V_0$ & $R_0$ & $a_0$ & $V_{LS}$ & $R_{LS}$ & $a_{LS}$ & $R_c$ \\
\hline
$^{90}$Zr & $\pi$ & 55.88 & 5.69 & 0.73  &  7.70  & 5.68 & 0.73  & 6.40 \\
          & $\nu$ & 48.12 & 5.69 & 0.73  &  7.70  & 5.68 & 0.73  &  \\
\hline
$^{132}$Sn & $\pi$ & 58.85 & 6.40  & 0.70  &  8.95  & 6.40 & 0.70  & 6.40 \\
           & $\nu$ & 47.50 & 6.10  & 0.70  &  5.50  & 6.10 & 0.70  &  \\
\hline
\end {tabular}
\end{center}
\vspace*{-0.5cm}
\caption{\small Parameters of the Woods-Saxon potential for the 
  $^{90}$Zr and
  $^{132}$Sn nuclei. The values of $V_0$ and $V_{LS}$  are expressed  
  in MeV, all the others 
  in fm. As in the traditional nuclear structure convention,
  we indicate with $\pi$ and $\nu$ the proton and  neutron parameters, 
  respectively. The explicit expression of the Woods-Saxon potential
  is given in Ref. \cite{ari07}.
}
\label{tab:WS}
\vspace*{-0.5cm}
\end{table}

The calculations have been done in discrete s.p. basis. The s.p.
Schr\"odinger equation with the Woods-Saxon potential has
been solved by expanding the s.p.  wave functions in a harmonic
oscillator basis. This produces bound states even for positive
s.p. energies. We consider configuration spaces by 
6 major
oscillator shells for oxygen, of 8 shells for calcium and 11
shells for zirconium, tin and lead nuclei. In our phenomenological
model the truncation of the s.p. configuration space is taken into
account in an effective manner by the choice of the parameters of the
effective interaction.

Our RPA calculations have been done with a zero range force of 
Landau-Midgal type,
\beqn
V_{\rm eff}(1,2)&=&
     \Big[  v_1(r_{12}) \, 
  + \, v_1^\rho(r_{12})\, \rho(r_1,r_2) \, + \, \left[ v_2(r_{12}) \, 
  + \, v_2^\rho(r_{12}) \, \rho(r_1,r_2) \right] \, \t1t2\nonumber \\
&&+ \, v_3(r_{12})\, \s1s2 \, 
  + \, v_4(r_{12}) \, \s1s2 \, \t1t2  
\Big] \delta(r_{12})
\, .
\label{eq:int}
\eeqn
In the equation above we have indicated with $\bsigma$ and $\btau$ the
usual Pauli spin and isospin operators. The zero range character of
the force implies that the $v_\alpha(r_{12})$ functions of the 
expression above are constants.  We used the values of the
constants defined in Ref. \cite{don09}. The choice of the parameters 
was done in two steps. First, we defined the values
for the density independent terms of the expression
(\ref{eq:int}). These values were fixed once for all the nuclei
and they describe the properties of some specific magnetic excitations
in $^{16}$O and $^{208}$Pb.  In MeV fm$^3$ units these values are
\beq
v_1\, = \, -918\,; \,\, v_2\,=\, 600\,; \,\, v_3 \, = \, 20\,;\,\,
v_4\,=\,200\,\,\,\,\,\,\, ,
\eeq
where the sub-indexes refer to the terms in the expression 
(\ref{eq:int}).
In the second step we chose the parameters of the density dependent
terms in order to reproduce the energies of the collective low-lying
3$^-$ states in $^{16}$O, $^{40}$Ca and $^{208}$Pb nuclei.  This
procedure selects the values of $v^\rho_1$. 
The parameters 
of the isospin dependent terms $v^\rho_2$ were chosen to reproduce 
the centroid energies of the GDR in $^{16}$O, $^{40}$Ca, $^{132}$Sn and
$^{208}$Pb. For each doubly closed shell nucleus considered,  
we give in Table \ref{tab:force} the values of the parameters of the
density dependent terms of the force.  Note that for the $^{90}$Zr
nucleus we used the set of values selected for $^{40}$Ca.

%
\begin{table}[hb]
\begin{center}
\begin{tabular}{cccccc}
\hline
   & $^{16}$O & $^{40}$Ca  & $^{90}$Zr   & $^{132}$Sn & $^{208}$Pb \\
\hline
 $v_1^\rho$ & 436.4 & 492.3 & 492.3 &  585.0  & 599.0  \\
 $v_2^\rho$ &-310.0 &-150.0 &-150.0 &  -50.0  & 0.0 \\
\hline
\end {tabular}
\end{center}
\vspace*{-0.5cm}
\caption{\small Values of the parameters of the density dependent
  terms of the interaction Eq. (\ref{eq:int}), in MeV fm$^3$. 
}
\label{tab:force}
\vspace*{-0.3cm}
\end{table}

The structure of the interaction (\ref{eq:int}) is simple if compared
with the complexity of modern microscopic nucleon-nucleon
interactions, as for example the Argonne V18 \cite{wir95}. We 
tested the reliability of our results by doing calculations also with
more elaborated effective nucleon-nucleon interactions. We used
the finite range interactions of Refs. \cite{don08t,don09}
containing also tensor terms. In the excitation of the 1$^-$ states,
the differences between the results obtained with the various
interactions are rather small and not relevant for the purposes of the
present work. For this reason we present here only the results
obtained with the interaction (\ref{eq:int}).

For a given multipolarity $J^\pi$, our RPA calculations produce a
number of solutions equal to the number of particle-hole excitations,
$N_{ph}$, compatible with the angular momentum and parity conservation
rules within the given configuration space.  For a single solution,
of excitation energy $\omega$, the RPA provides the set of amplitudes
$X_{ph}(\omega)$ and $Y_{ph}(\omega)$ which describe the wave function
of the excited state in terms of particle-hole (p-h) and hole-particle
(h-p) excitations, respectively. The proper normalization of the
many-body wave function implies that, for a given excited state, the
RPA amplitudes are normalized as:
\beq
\sum_{ph=1}^{N_{ph}} 
\left[ X^2_{ph}(\omega) - Y^2_{ph}(\omega) \right] = 1
\,\,\,.
\label{eq:norm}
\eeq
In our search for states which can be identified as PDR, we 
singled out a few quantities which summarize the main characteristics
of each state.  First, for a given excited state, we calculated
the relative contribution of protons and neutrons to the normalization
(\ref{eq:norm}).  These contributions, indicated as \npi and \nnu in
the following, are obtained from Eq.~(\ref{eq:norm}) by summing
over p-h pairs for protons, or, respectively, neutrons only.

Second, we defined an index to measure the degree of collectivity
of a specific excited state. In the ideal
collective state all the p-h excitation pairs
contribute with the same statistical weight. In this case, all
the $X^2_{ph}(\omega) - Y^2_{ph}(\omega)$ terms of Eq. (\ref{eq:norm})
would contribute $1/N_{ph}$. From these considerations we defined a
collectivity index as 
\beq
\doc = N^* / N_{ph}  
\,\,\,,
\label{eq:doc}
\eeq
where $N^*$ is the number of states with 
$(X^2_{ph}(\omega) - Y^2_{ph}(\omega)) \ge 1/N_{ph}$. 
The two extreme values of $\doc$ are 1 in the fully collective
case, and $1/N_{ph}$ when the excitation is produced by a single p-h
pair.

The definition of $\doc$, Eq.~(\ref{eq:doc}), depends on the number of
p-h excitations $N_{ph}$, and this latter quantity is related to the
size of the configuration space. The values of the index $\doc$ must
be used to compare excited states calculated within the same
configuration space. To gauge the values of $\doc$ indicating a high
degree of collectivity, we calculated $\doc$ for the collective
low-lying $3^-$ states of various doubly closed shell nuclei.  These
values are given in Table \ref{tab:3-} and are our reference values.

%
\begin{table}[h]
\begin{center}
\begin{tabular}{ccccc}
\hline
   & $^{16}$O & $^{40}$Ca  & $^{132}$Sn & $^{208}$Pb \\
\hline
 $\omega$  & 6.12    & 3.74   &  4.34 & 2.63\\
 \nst      & 5       & 10     &  27   & 40\\
 $\doc$    & 0.192 & 0.135 &  0.095 & 0.119 \\
 \npi      & 0.501 & 0.568 &  0.187 & 0.362 \\
 \nnu      & 0.499 & 0.432 &  0.813 & 0.638 \\
\hline
\end {tabular}
\end{center}
\vspace*{-0.5cm}
\caption{\small Values of the collectivity indexes for the low-lying
  $3^-$ states of the doubly closed shell nuclei we 
  considered. The values of the excitation energies $\omega$ are
  expressed in MeV, 
  $\doc$ is defined in Eq. (\ref{eq:doc}) and  \nst is the numerator
  of that equation. With \npi and \nnu we have indicated
  respectively the proton and neutron contribution to the
  normalization (\ref{eq:norm}), clearly \npi+\nnu=1.
}
\label{tab:3-}
\vspace*{-0.35cm}
\end{table}

The collectivity of a state is not only related to the value of the
$X$ and $Y$ amplitudes, but also to the coherence of the p-h pairs in
constructing transition amplitudes. For this reason we also
calculated the transition densities 
\beq
\rho(EJ;\omega,r)
= \sum_{ph} \left[ X_{ph}(\omega) + Y_{ph}(\omega)\right]
\rho^J_{ph}(r)
\,\,\,,
\label{eq:trans1}
\eeq
with
\beq
\rho^J_{ph}(r) =  (-1)^{j_p + \half} \,\,
\frac {\what{j_p}\,\what{J}\,\what{j_h}}{\sqrt{4 \pi}}
\threej {j_p}{J}{j_h}{\half}{0}{-\half}
\,R_p(r)R_h(r)
\,\,\,.
\label{eq:trans2}
\eeq
In the equation above we have indicated with $R(r)$ the radial part 
of the s.p. wave functions, with $j$ their angular
momenta, we used the symbol
$\what{j}=\sqrt{2j+1}$ and the traditional symbol to indicate the
Wigner 3-j coefficient.

\begin{figure}[ht]
\begin{center}
\includegraphics[scale=0.5, angle=0] {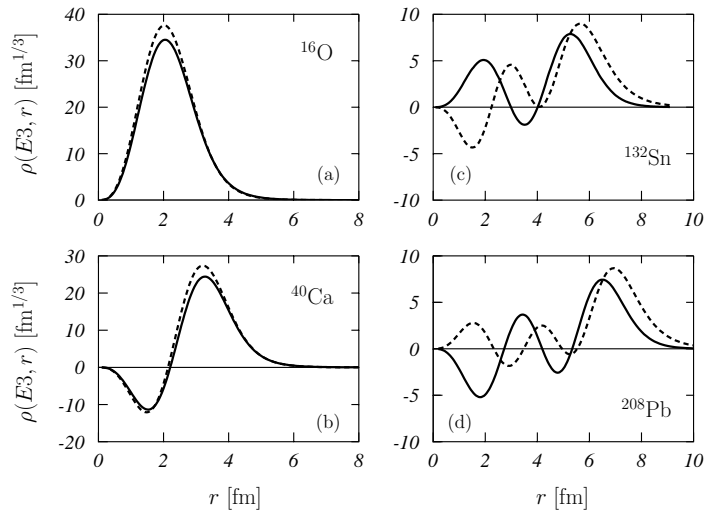} 
\vspace*{-0.3cm}
\caption{\small Transition densities for the 3$^-$ low-lying states of
  the $^{16}$O, $^{40}$Ca, $^{132}$Sn and $^{208}$Pb nuclei.  The full
  lines indicates the proton transition densities, and the dashed
  lines the neutron transition densities. Here, and in the following
  figures, we drop the dependence on $\omega$ with respect to the
  definition (\ref{eq:trans1}), since each transition density is
  calculated for a specific value of the excitation energy.
  }
\label{fig:3-}
\end{center}
\vspace*{-1cm}
\end{figure}

We calculated separately the proton and neutron transition densities
by limiting the sum of Eq. (\ref{eq:trans1}) to proton or neutron
pairs only. The transitions densities for the low-lying 3$^-$ states
of the various doubly closed shell nuclei are shown in
Fig. \ref{fig:3-}. The full lines indicate the proton densities, while
the dashed lines the neutron densities. The structure of the various
densities becomes more complicated the heavier is the nucleus. 
Despite these differences, in all the cases shown in
the figure the {\sl in phase} behavior of the two
types of density is evident.
This indicates the $IS$ character of the
transition. The $IS$ structure of these states is confirmed by the fact
that the energies eigenvalues are sensitive only to scalar terms of
the interaction, $v_1$ and $v^\rho_1$ of Eq. (\ref{eq:int}).

Finally, we characterize each state by its $B(E1)$ value and by the
ratio $\rbe$ between the $B(E1)$ value of the specific state and the
total $B(E1)$ strength. We prefer to consider this ratio, rather than
making a comparison with the Thomas-Reiche-Khun energy weighted sum
rule, because our approach is not self-consistent, and in addition it
uses a truncated s.p. configuration space. In any case,
our results satisfy the sum rule at the 5\% level.

In our work, we used the following strategy.  The set of s.p. wave
functions, and the parameters of the effective nucleon-nucleon
interaction, were chosen to reproduce some properties of the $^{16}$O,
$^{40}$Ca, $^{90}$Zr, $^{132}$Sn and $^{208}$Pb doubly magic nuclei,
as we have discussed above. Around each doubly magic nucleus we
constructed a set of isotopes by increasing or decreasing, within the
chosen configuration space, the number of neutron levels forming the
ground state.  Since we work with a spherical basis, the difference
between the number of neutrons of each isotope is $2j+1$, where $j$ is
the angular momentum of the level with higher energy. We considered
only isotopes which have been experimentally identified.  For each
isotopic chain we used the effective interaction and s.p. basis
adjusted to reproduce the properties of the doubly magic nucleus of
the chain. The number of neutrons was changed by considering a
different number of fully occupied s.p. levels.

Before presenting the results of our calculations we want to
critically discuss the basic features and the limits of our model.
The first point is related to the choice of a discrete, and
restricted, configuration space.  We have recently verified the large
sensitivity of the RPA results to the truncation of the configuration
space \cite{don08t,don09}. Only a proper treatment of the continuum
can provide numerically stable RPA results. This is a big problem in
self-consistent calculations where the effective nucleon-nucleon
interaction used in the RPA is the same one  
also used to build the s.p.  basis by
means of a Hartree-Fock calculation. In our phenomenological approach
we use s.p. bases constructed on a Woods-Saxon potential, and
effective interactions chosen to reproduce the energies of some specific
excited states.  The effects of the truncation of the configuration
space are effectively taken into account by the choice of the parameters
of the interaction.  Therefore our effective interactions are strictly
related to the configuration space.  In our calculations all the
nuclei of a given isotopic chain are described by using the same
parameterization of the nucleon-nucleon interaction and the same set
of s.p.  wave functions. This ensures numerical stability at the price
of a rigid use of the s.p. wavefunctions.  A Hartree-Fock approach
would be more flexible.  In any case, we studied the ground states of
the oxygen isotopes $^{16}$O, $^{22}$O, $^{24}$O, $^{28}$O by using a
spherical Hartree-Fock approach \cite{bau99,don09} with the Gogny D1
interaction \cite{dec80}, and we did not find relevant differences in
the occupied s.p. wave functions of the $^{16}$O core.

The second point we would like to discuss is related to the fact that
our calculations do not consider effects beyond one-particle one-hole
(1p-1h) excitations, even though in the literature there are now quite
a few
calculations of the PDR excitations where these effects are taken into
account \cite{rye02,adr05,har04,tso04a,tso04b,tso06,sar04,lit07}.  The
inclusion of p-h excitations beyond those considered by the RPA
produces two effects related to the real and imaginary part of the
self-energy. The real part of the self-energy changes the position of
the resonance. In our approach this effect is taken into account by
using phenomenological s.p. energies and effective interactions.  The
imaginary part generates a spreading of the width of the resonances
obtained in RPA. Our approach cannot simulate this effect.

Finally, we consider pairing and deformation. Our phenomenological
approach cannot simulate these effects. For this reason, besides
doubly magic nuclei, we have studied only those nuclei with fully
occupied s.p. levels. In this case, we expect that the spherical
symmetry of the nucleus is almost restored, and also pairing effects
should be smaller than in nuclei with the partially occupied levels. 

An accurate description of the experimental data requires the
inclusion of terms which consider the spreading of the resonance
width. However, despite of its simplicity, our model should
predict the position of the resonance, its total strength, the
degree of collectivity and the relative importance of proton and
neutron excitations. These are the quantities we have considered in
our work and they are presented and discussed in the next section.

\section{Specific applications}
\label{sec:resu}

In this section we present the results obtained by applying our model
to a set of isotopic chains built around the doubly magic nuclei
$^{16}$O, $^{40}$Ca, $^{90}$Zr and $^{132}$Sn.  Before doing that, we
discuss the $^{208}$Pb results. In this nucleus we identify the PDR,
therefore the values of the collectivity indexes, and the behavior of
the proton and neutron transition densities, can be used as
references for the results obtained for the other nuclei.

\begin{figure}[th]
\begin{center}
\includegraphics[scale=0.5, angle=0] {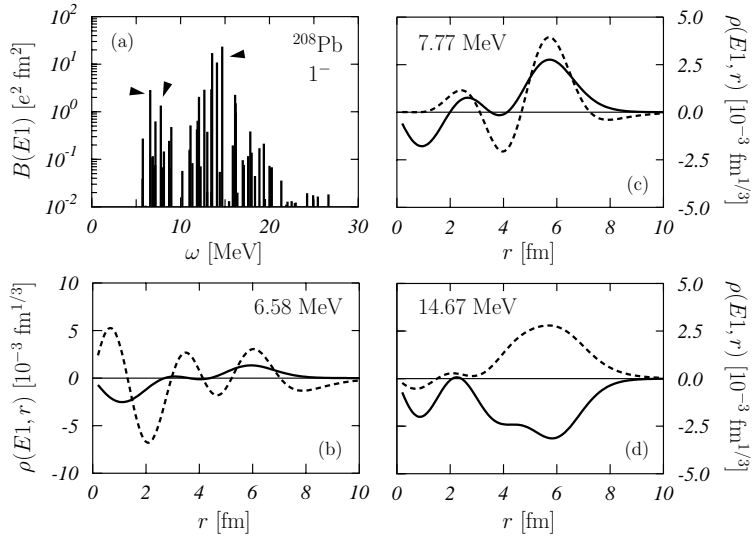} 
\vspace*{-0.3cm}
\caption{\small Dipole results for the $^{208}$Pb nucleus.  In panel
(a) we show the $B(E1)$ values as a function of the excitation
energy. In the other panels, we present the transition densities for
the states indicated by the arrows, whose excitation energies are
given in the panels.  The meaning of the lines in panels (b), (c) and
(d) is the same as in Fig. \ref{fig:3-}.  
}
\label{fig:pb1-}
\end{center}
\vspace*{-0.7cm}
\end{figure}

In panel (a) of Fig. \ref{fig:pb1-} we present the $B(E1)$ values of the
$^{208}$Pb nucleus as a function of the excitation energy $\omega$.
The figure shows the GDR region which has its maximum at 14.67 MeV and
a set of peaks at lower energy which we identify with the PDR region.
We discuss with some detail the characteristics of the states which
are indicated by the arrows in panel (a). The energy value of the
first state is 6.58 MeV. For this state we found $\doc$=0.049 and
\nst=10. The proton contribution to the normalization of the
wavefunction is \npi=0.026, therefore the neutron contribution is
\nnu=0.974. The proton and neutron transition densities for this state
are shown in panel (b) of the figure. The characteristics of the other
state are $\omega$=7.77 MeV, $\doc$=0.049, \nst=10, \npi=0.238 and
\nnu=0.762, and its transition densities are shown in panel (c) of the
figure.  For both states, the proton and neutron transition densities
are {\sl in phase}, which is the typical behavior of the $IS$
excitation.  The behavior of these transition densities, are rather
different from those of the states forming the GDR.  As typical
example, we discuss here only the state at 14.67 MeV.  For this state
we obtain $\doc$=0.073, and \nst=15, two values indicating a slightly
higher collectivity with respect to that of the PDR. The contributions
to the normalization of the wave functions are \npi=0.682 and
\nnu=0.318.  The transition densities for the GDR state are shown in
panel (d) and they have an {\sl out of phase} behavior indicating the
$IV$ nature of the excitation.

Photon scattering experiments on $^{208}$Pb have identified, in
addition to the well known GDR peaked at 13.5 MeV \cite{ahr75}, a
small dipole resonance around 7.35 MeV which has been interpreted as
PDR \cite{rye02}.  In our calculations we found a large resonance
above 11.0 MeV which contributes for about the 70\% of the total
dipole strength, and a tiny resonance with centroid energy at about
7.2 MeV which carries about the 5\% of the total strength.  The two
resonances have a similar degree of collectivity, though the resonance
at smaller energy shows an $IS$ structure while the other resonance
has an $IV$ structure.  In the PDR, the contribution of neutron p-h
pairs is slightly larger than that of the protons, while in the GDR
the neutron and proton contributions are more equilibrated.

In our phenomenological approach the force paramaters have been chosen
to reproduce the GDR centroid energy and the energy of the low lying
3$^-$ state. With this choice we have obtained a PDR whose 
centroid energy is about 6.2 MeV lower than that of the GDR, a
value close to the experimental one of 6.15
MeV \cite{ahr75,rye02}. A similar result is obtained in the
phenomenological approach of Ref. \cite{tse07}, where in addition to
the usual RPA degrees of freedom also more complex p-h excitations are
considered by using phonon coupling technique.

It is interesting to compare our results with those obtained by the
more ambitious self-consistent calculations in both relativistic
\cite{vre01b,lit07} and non-relativistic \cite{sar04,lyu08}
frameworks. Since in self-consistent calculations the interaction
parameters are chosen mainly to reproduce the ground state properties
of nuclei, the position of the GDR centroid energy is a prediction of
the theory. For this reason we think that the comparison with
experimental GDR and PDR centroid energies is not really illuminating,
because of the many details entering in the calculations. Instead, we
believe it is more useful to compare the differences between GDR and
PDR centroid energies obtained in the various calculations.  This
difference is essentially related to the relative strength of the $IS$
and $IV$ components of the effective interaction.  The relativistic
RPA calculations of Ref. \cite{vre01b} predict a difference of about
5.7 MeV, similar to the 5.9 MeV obtained in the relativistic RPA
calculation of Ref. \cite{lit07}. In this reference it is shown that
the inclusion of higher order p-h excitations, by means of phonon
coupling procedure, produces a small lowering of the GDR centroid
energy, 0.2 MeV, but a larger shift of the PDR centroid energy, and
the difference between these two quantities is 6.6 MeV, close to the
experimental value. The results of the non-relativistic calculations
of Ref. \cite{sar04} complicate the picture. The inclusion of the
phonon coupling lowers the RPA GDR centroid energy by 0.5 MeV thus
showing an effect which is qualitatively and quantitatively similar to
that found in relativistic calculations. On the contrary, the effect
of the phonon coupling on the energy difference has opposite sign.
The RPA calculations generate an energy difference of 6.9 MeV which
becomes smaller, 5.8 MeV, after the inclusion of the phonon coupling.
The non-relativistic calculations of Ref. \cite{lyu08}, which consider
the p-h excitations in the continuum, show that the inclusion of the
phonon coupling lowers the GDR centroid energy of about 0.6 MeV,
confirming the general trend. Unfortunately in this article the
presence of the PDR is not discussed, even though the photoabsorption
cross section shows a tiny resonance at about 8 MeV, only 3 MeV below
the peak of the GDR.  The situation is not yet clear and deserves
further investigation.

In the following study, we use the values of the indexes $\doc$, \npi
and \nnu presented above as a guide to identify the presence of a
collective excitation, and we identify the $IS$ or $IV$ character of the
excitation by analyzing the behavior of the proton and neutron
transition densities. 

\begin{figure}[th]
\begin{center}
\includegraphics[scale=0.5, angle=0] {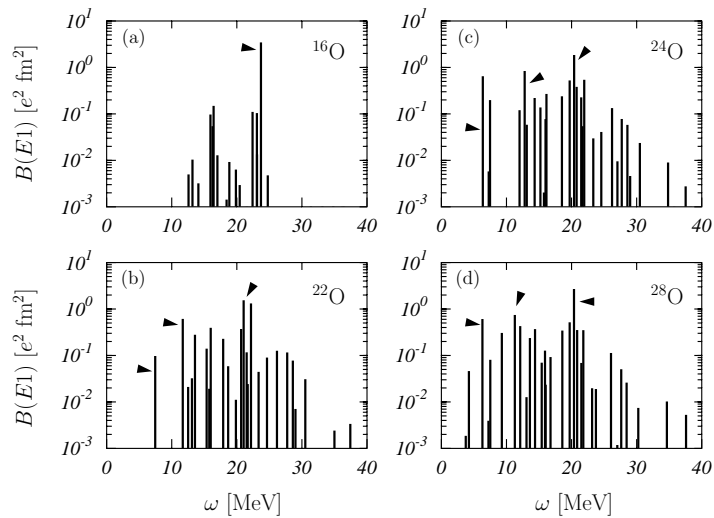} 
\vspace*{-0.3cm}
\caption{\small $B(E1)$ distributions for the oxygen isotopes we have
  studied. The collectivity indexes of the states indicated by the arrows
  are given in Table \ref{tab:oxy}.
  }
\label{fig:oxy}
\end{center}
\vspace*{-0.7cm}
\end{figure}

%
\begin{table}[h]
\begin{center}
\begin{tabular}{ccccccc}
\hline
 nucleus & $\omega$ [MeV] & \nst & $\doc$ & \npi & \nnu & $\rbe$ \\
\hline
 $^{16}$O & 23.70   & 4      &0.143 & 0.480 & 0.520 & 0.855 \\
\hline
 $^{22}$O & 7.46   & 1      &0.003 & 0.006 & 0.074 & 0.017 \\
          &11.68   & 5      &0.152 & 0.074 & 0.926 & 0.106 \\
          &22.17   & 5      &0.152 & 0.459 & 0.541 & 0.229 \\
\hline
 $^{24}$O & 6.36   & 2      &0.057 & 0.020 & 0.980 & 0.095 \\
          &12.01   & 6      &0.171 & 0.546 & 0.454 & 0.018 \\
          &20.40   & 9      &0.257 & 0.530 & 0.490 & 0.273 \\
\hline
 $^{28}$O & 6.29   & 3      &0.077 & 0.035 & 0.965 & 0.079 \\
          &11.28   & 8      &0.205 & 0.162 & 0.838 & 0.096 \\
          &20.36   & 10     &0.256 & 0.634 & 0.366 & 0.348 \\
\hline
\end {tabular}
\end{center}
\vspace*{-0.5cm}
\caption{\small Values of the collectivity indexes for the
  $1^-$ states of the various oxygen isotopes identified by the arrows in
  Fig. \ref{fig:oxy}. The meaning of the various indexes is the same
  as in Table \ref{tab:3-}. We also show the ratio $\rbe$ 
  between the $B(E1)$ of the specific state and the total $B(E1)$ strength.
}
\label{tab:oxy}
\vspace*{-0.2cm}
\end{table}

\begin{figure}[ht]
\begin{center}
\includegraphics[scale=0.45, angle=0] {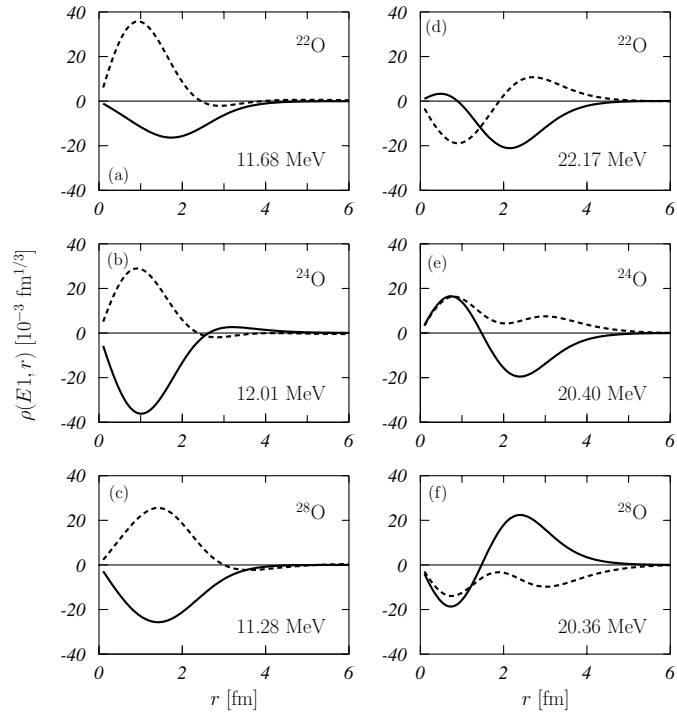} 
\vspace*{-0.3cm}
\caption{\small Transition densities of some 1$^-$ states for various
    oxygen isotopes. The numbers in the panels 
    indicate the excitation energy.
    Full and dashes lines represent proton and neutron densities,
    respectively. 
    }
\label{fig:toxy}
\end{center}
\vspace*{-0.7cm}
\end{figure}

We start our discussion by considering the $^{22}$O, $^{24}$O,
$^{28}$O nuclei, obtained from the $^{16}$O core by filling the
neutron $1d_{5/2}$, $2s_{1/2}$ and $1d_{3/2}$ s.p. levels
respectively. The $B(E1)$ distributions for these isotopes are presented
in Fig. \ref{fig:oxy}.  The isotopes with neutron excess show a rich
structure at excitation energies below the GDR peak.  The values of
the collectivity indexes of the states indicated by the arrows in the
figure are given in Table \ref{tab:oxy}. In this table we also give the
ratio $\rbe$ between the $B(E1)$ of the indicated state and the global
$B(E1)$ strength.

The values presented in Table \ref{tab:oxy} give some indication of
collectivity of the states below the GDR.  In the three heavier
isotopes of the $^{16}$O, the states around 11, 12 MeV have a
relatively large degree of collectivity. For this reason we analyzed
their transition densities, and we found that they have an $IV$
structure, as it is shown in Fig. \ref{fig:toxy}.  For these three
nuclei we compare the transition densities of the GDR peak with that
of the state with the largest $B(E1)$ value. The {\sl out of phase}
behavior of the densities indicates that these states are produced by
the fragmentation of the GDR and they are not a new type of
excitation.

\begin{figure}[ht]
\begin{center}
\includegraphics[scale=0.5, angle=90] {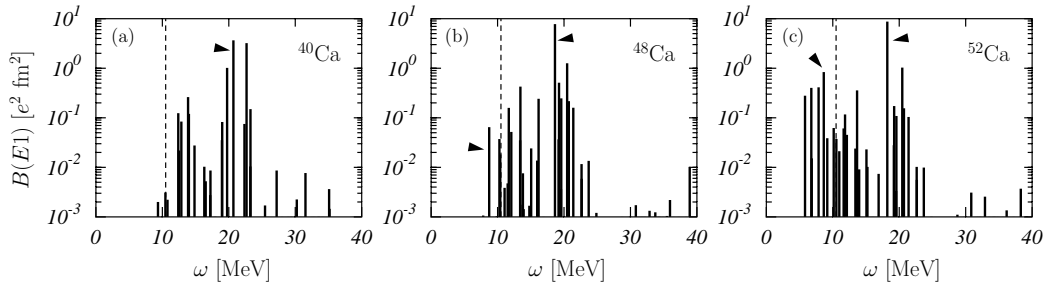} 
\vspace*{-0.3cm}
\caption{\small Same as in Fig. \ref{fig:oxy},
  but for the calcium isotopes we have
  studied. The collectivity indexes of the states indicated by arrows
  are given in Table \ref{tab:ca}.
  The meaning of the dashed lines is explained in Fig.~\ref{fig:int}
  and in the related text.
  }
\label{fig:ca}
\end{center}
\vspace*{-0.5cm}
\end{figure}

\begin{figure}[ht]
\begin{center}
\includegraphics[scale=0.5, angle=0] {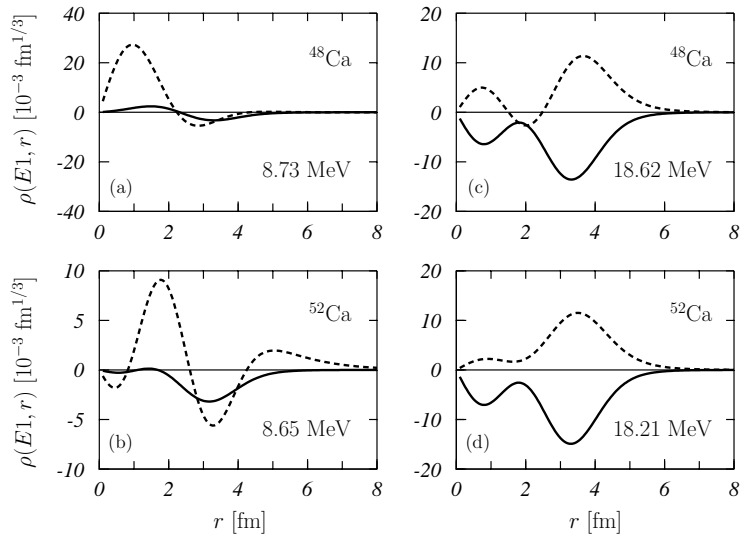} 
\vspace*{-0.3cm}
\caption{\small Same as in Fig.\ref{fig:toxy}, but 
for various calcium isotopes. 
    }
\label{fig:tca}
\end{center}
\vspace*{-0.7cm}
\end{figure}

The situation is rather different for the other isotopic chains we have
studied. The results for the calcium chain are presented in
Figs. \ref{fig:ca} and \ref{fig:tca} and in Table \ref{tab:ca}. In our
study, we considered the $^{48}$Ca and $^{52}$Ca isotopes obtained
from the $^{40}$Ca core by filling the neutron $1f_{7/2}$ and
$2p_{3/2}$ s.p. levels. The $B(E1)$ distributions of these three
isotopes are shown in Fig. \ref{fig:ca}. As we have already observed
for the oxygen case, also in this case the presence of excess neutrons
produces $E1$ strength at energies lower than those of the GDR. We have
repeated the study of these states in analogy to what we have done for
lead and oxygen, and the values of the collectivity indexes for the
states indicated by the arrows are given in Table \ref{tab:ca}. The
states at the peak of the GDR show a degree of collectivity comparable
with that of the 3$^-$ state (see Table \ref{tab:3-}).  Proton and
neutron p-h pairs contribute to the excitation with almost equal
weight.  The structure of the lower energy states is different. There
is a certain degree of collectivity, but it is smaller than that of
the states just described. In addition it is evident that these states
are dominated by neutron excitations. The different structure of these
two type of states is emphasized by the transition densities, which
are shown in Fig. \ref{fig:tca}. The low energy states show an {\sl in
phase} behavior, typical of the $IS$ transition, while the higher
energy states have the typical $IV$ {\sl out of phase} behavior. All
these remarks indicate that the low energy states are not produced by
the fragmentation of the GDR, as it happens for the oxygen isotopes,
but they are a new type of excitation.

%
\begin{table}[hb]
\begin{center}
\begin{tabular}{ccccccc}
\hline
  nucleus & $\omega$ [MeV]& \nst & $\doc$ & \npi & \nnu & $\rbe$ \\
\hline
 $^{40}$Ca &  20.69   & 8      &0.133 & 0.670 & 0.340 & 0.409 \\
\hline
 $^{48}$Ca &   8.73    & 5      &0.077 & 0.096 & 0.904 & 0.006 \\
           &  18.62    & 11     &0.170 & 0.521 & 0.479 & 0.684 \\
\hline
 $^{52}$Ca &   8.65    & 3      &0.045 & 0.020 & 0.980 & 0.063 \\
           &  18.21    & 9      &0.134 & 0.503 & 0.497 & 0.664 \\
\hline
\end {tabular}
\end{center}
\vspace*{-0.5cm}
\caption{\small Same as in Table \ref{tab:oxy}
  for the
  $1^-$ states of the various calcium isotopes identified by the arrows in
  Fig. \ref{fig:ca}. 
}
\label{tab:ca}
\vspace*{-0.3cm}
\end{table}

\begin{figure}[ht]
\begin{center}
\includegraphics[scale=0.45, angle=0] {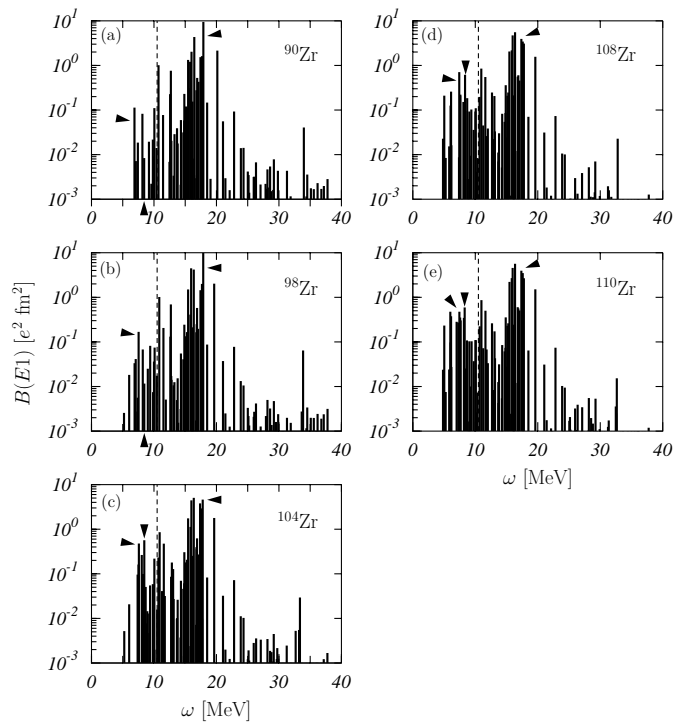} 
\vspace*{-0.3cm}
\caption{\small Same as in Fig. \ref{fig:oxy}, but 
  for the zirconium isotopes we have
  studied. 
  The collectivity indexes of the states indicated by arrows
  are given in Table \ref{tab:zr}.
  }
\label{fig:zr}
\end{center}
\vspace*{-0.8cm}
\end{figure}

When the nucleon number increases, the features related to the rising
of the PDR become more evident. We found an excellent example of this
fact in the Zr chain.  In Fig. \ref{fig:zr} we show the $B(E1)$
distributions for the $^{90}$Zr, $^{98}$Zr, $^{104}$Zr, $^{108}$Zr,
and $^{110}$Zr isotopes. We constructed this isotopic chain starting
from the $^{90}$Zr core and filling the neutron $1g_{7/2}$,
$2d_{5/2}$, $2d_{3/2}$ and $3s_{1/2}$ s.p. levels respectively. In
these calculations we used the same interaction adopted for the Ca
calculations.

%
\begin{table}[hb]
\begin{center}
\begin{tabular}{ccccccc}
\hline
 nucleus & $\omega$ [MeV]& \nst & $\doc$ & \npi & \nnu & $\rbe$ \\
\hline
 $^{90}$Zr &  7.41    & 4      &0.026 & 0.022 & 0.978 & 0.000 \\
           &  8.42    & 7      &0.045 & 0.024 & 0.976 & 0.000 \\
          &  17.89    & 19     &0.123 & 0.656 & 0.344 & 0.316 \\
\hline
 $^{98}$Zr &  7.53    & 8      &0.050 & 0.013 & 0.987 & 0.005 \\
           &  8.45    & 7      &0.043 & 0.019 & 0.981 & 0.000 \\
           & 17.89    & 18     &0.111 & 0.678 & 0.322 & 0.313 \\
\hline
 $^{104}$Zr & 7.52    & 6      &0.036 & 0.011 & 0.989 & 0.004 \\
           &  8.45    & 11     &0.066 & 0.025 & 0.975 & 0.016 \\
           & 17.79    & 13     &0.078 & 0.818 & 0.182 & 0.133 \\
\hline
 $^{108}$Zr & 7.57    & 6      &0.035 & 0.011 & 0.989 & 0.005 \\
           &  8.35    & 11     &0.065 & 0.025 & 0.975 & 0.017 \\
           & 17.58    & 10     &0.059 & 0.251 & 0.749 & 0.093 \\
\hline
 $^{110}$Zr & 7.45    & 7      &0.041 & 0.004 & 0.996 & 0.012 \\
           &  8.29    & 13     &0.076 & 0.020 & 0.980 & 0.016 \\
           & 17.58    & 10     &0.058 & 0.250 & 0.750 & 0.092 \\
\hline
\end {tabular}
\end{center}
\vspace*{-0.5cm}
\caption{\small Same as in Table \ref{tab:oxy}
   for the
  $1^-$ states of the various zirconium isotopes identified by the arrows in
  Fig. \ref{fig:zr}. 
}
\label{tab:zr}
\vspace*{-0.3cm}
\end{table}

With the increasing of the number of neutrons, we observe a
fragmentation of the $B(E1)$ distributions. There is a fragmentation of
the GDR and a rising of new strength at low energy. In $^{90}$Zr the
most important state is at 17.89 MeV, and is indicated by the arrow in
Fig. \ref{fig:zr}. This state carries about the 30\% of the total
$B(E1)$ strength.  In the $^{110}$Zr nucleus, this state is not any more
the most important one, and it is responsible only for about the 9\%
of the total strength. The remaining part of the GDR strength has been
redistributed between states with slightly smaller energies.

\begin{figure}[th]
\begin{center}
\includegraphics[scale=0.45, angle=0] {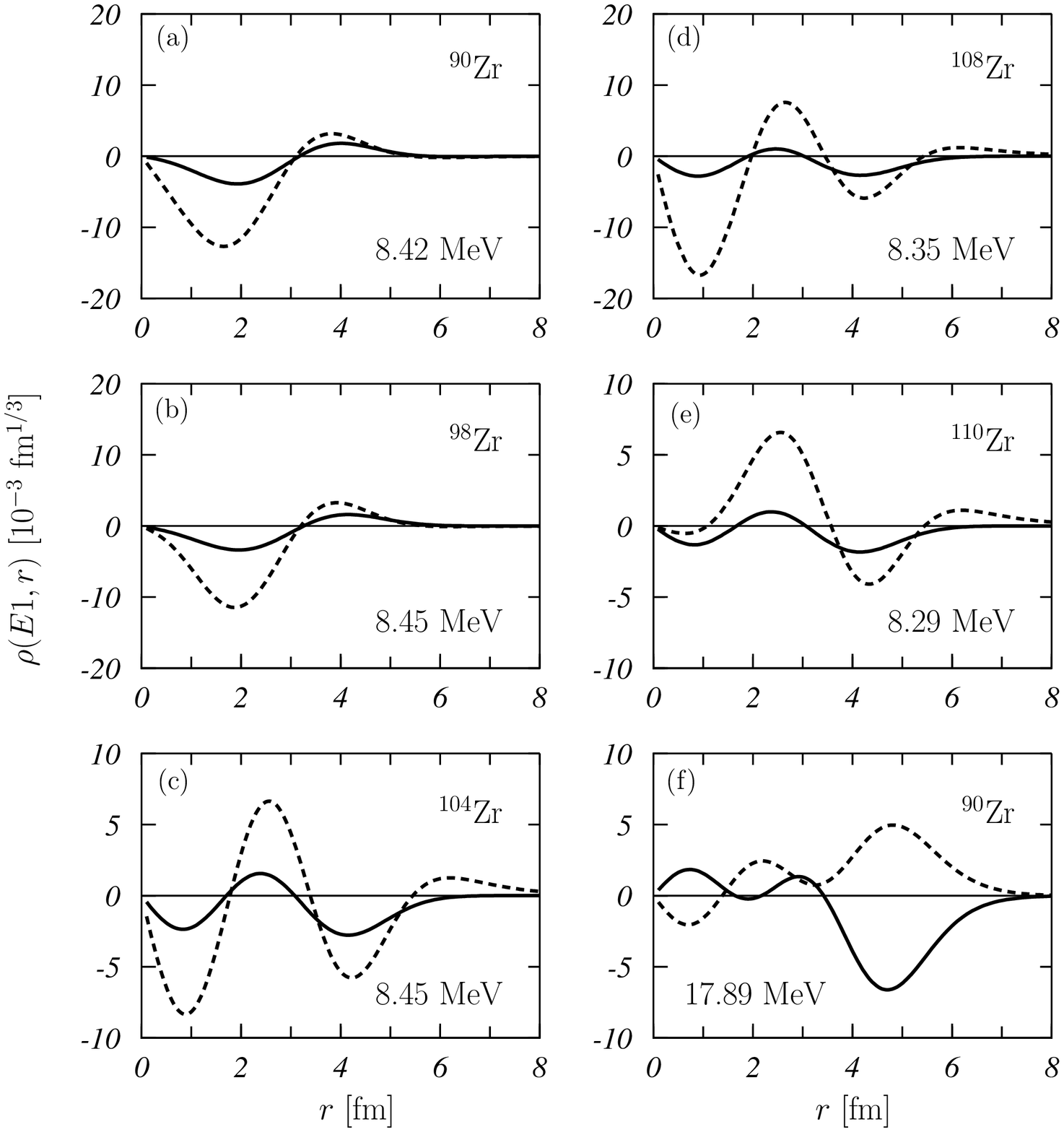} 
\vspace*{-0.3cm}
\caption{\small Same as in Fig.~\ref{fig:toxy}, but 
    for various
    zirconium isotopes. 
       }
\label{fig:tzr}
\end{center}
\vspace*{-0.7cm}
\end{figure}

\begin{figure}[bh]
\begin{center}
\includegraphics[scale=0.45, angle=0] {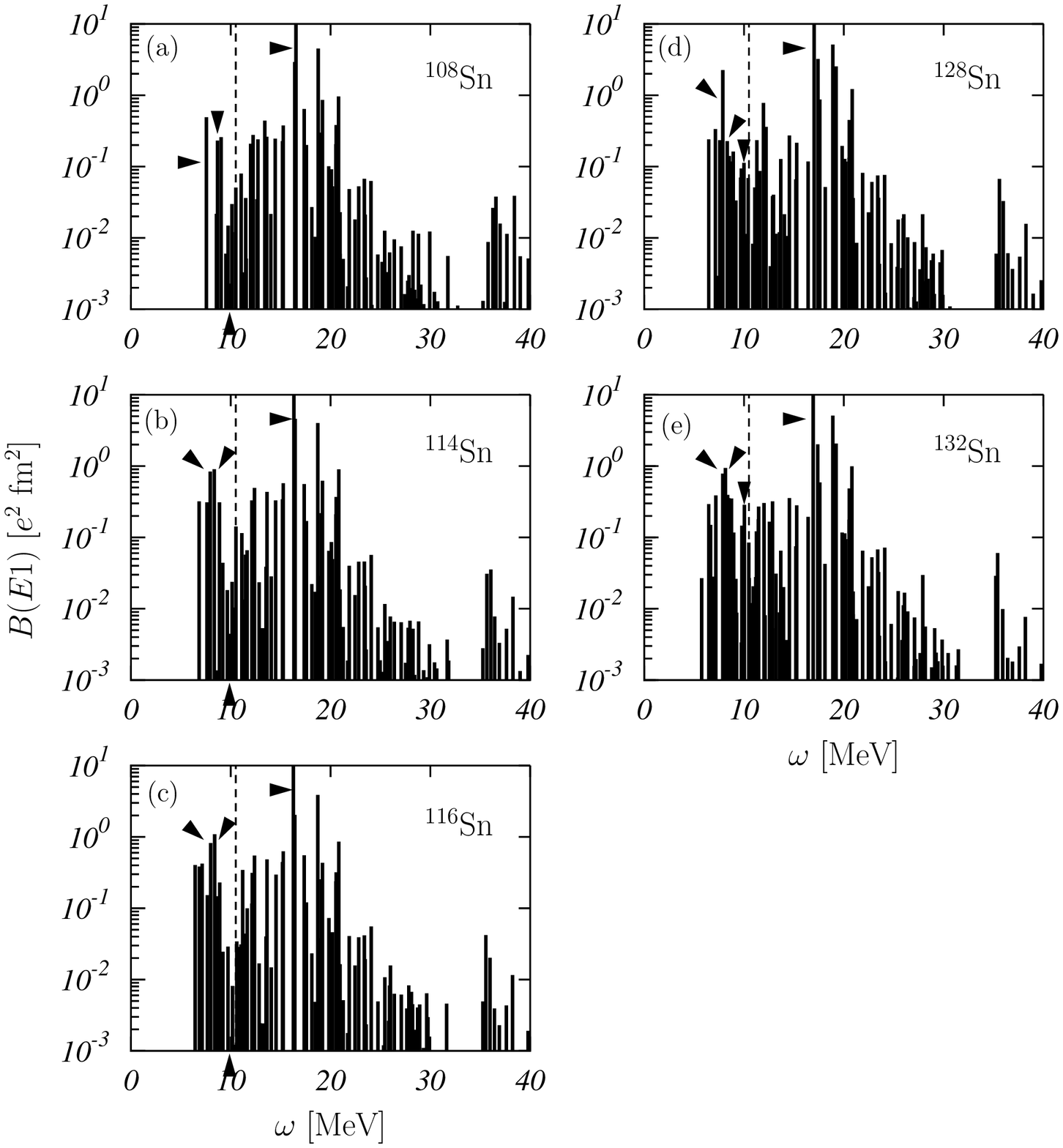} 
\vspace*{-0.3cm}
\caption{\small Same as in Fig.~\ref{fig:oxy}, but
  for the tin isotopes we have studied. 
  The collectivity indexes of the states indicated by arrows
  are given in Table \ref{tab:sn}.
  }
\label{fig:sn}
\end{center}
\vspace*{-0.7cm}
\end{figure}

In our discussion we consider three specific states for each isotope,
and we indicated them in Fig. \ref{fig:zr} by the arrows. The values
of the collectivity indexes for these states are presented in
Table \ref{tab:zr}. We have followed the evolution of these states in
all the isotopic chain. The lower energy states have extremely small
$B(E1)$ values in the $^{90}$Zr and $^{98}$Zr isotopes, as it is shown
by the values of $\rbe$. Their contributions to the total strength
increase with increasing the neutron number. Both states are neutron
dominated. On the contrary, the most important state of the GDR in
$^{90}$Zr has an almost equal contribution of proton and neutron
components. The proton contribution becomes smaller with the increase
of the neutron number, but still it remains remarkable.  The low-lying
states have an $IS$ structure, while the states of the giant resonance
have an $IV$ structure. As an example of the results we have obtained, we
show in Fig. \ref{fig:tzr} the transition densities for the 8.3 - 8.4
MeV states, and those of the peak in $^{90}$Zr at 17.89 MeV.  The
proton and neutron transition densities of the lower energy states
clearly show an $IS$ {\sl in phase} behavior, contrary to the {\sl out
of phase} behavior of the 17.89 MeV state in $^{90}$Zr indicating the
$IV$ character of the GDR.

As a last application of our model we show the results regarding a Sn
isotopic chain. These calculations are of interest since the PDR has
been recently identified around 10 MeV in tin isotopes \cite{adr05}.
In the case of the Sn chain, the doubly magic nucleus is the heaviest
of the chain, the $^{132}$Sn, which, obviously has the largest number
neutrons.  We obtain the other isotopes by removing neutrons from
the $2d_{3/2}$, $1h_{11/2}$, $3s_{1/2}$ and $2d_{5/2}$ levels, to
obtain the $^{128}$Sn, $^{116}$Sn, $^{114}$Sn and $^{108}$Sn nuclei, 
respectively.

%
\begin{table}[h]
\begin{center}
\begin{tabular}{ccccccc}
\hline
 nucleus & $\omega$ [MeV]& \nst & $\doc$ & \npi & \nnu & $\rbe$ \\
\hline
 $^{108}$Sn &  7.56    & 7      &0.048 & 0.132 & 0.868 & 0.014 \\
            &  8.67    & 13     &0.078 & 0.874 & 0.126 & 0.007 \\
            &   9.89    & 11     &0.066 & 0.058 & 0.942 & 0.000 \\
            &  16.53    & 21     &0.126 & 0.476 & 0.524 & 0.546 \\
\hline
 $^{114}$Sn &  7.95    & 9      &0.053 & 0.422 & 0.578 & 0.022 \\
            &   8.35    & 13     &0.076 & 0.583 & 0.417 & 0.024 \\
            &   9.88    & 6      &0.035 & 0.025 & 0.975 & 0.000 \\
            &  16.33    & 24     &0.140 & 0.448 & 0.552 & 0.511 \\
\hline
 $^{116}$Sn &  7.98    & 7      &0.040 & 0.498 & 0.502 & 0.021 \\
            &   8.38    & 14     &0.081 & 0.530 & 0.470 & 0.028 \\
            &   9.88    & 3      &0.017 & 0.011 & 0.989 & 0.000 \\
            &  16.27    & 24     &0.139 & 0.509 & 0.491 & 0.569 \\
\hline
 $^{128}$Sn &  7.88    & 9      &0.051 & 0.166 & 0.834 & 0.053 \\
            &   8.31    & 11     &0.062 & 0.805 & 0.195 & 0.005 \\
            &   9.99    & 7      &0.039 & 0.067 & 0.933 & 0.003 \\
            &  17.03    & 17     &0.096 & 0.425 & 0.575 & 0.450 \\
\hline
 $^{132}$Sn &  7.88    & 7      &0.038 & 0.038 & 0.962 & 0.018 \\
            &   8.13    & 12     &0.066 & 0.489 & 0.501 & 0.022 \\
            &  10.02    & 14     &0.077 & 0.130 & 0.870 & 0.007 \\
            &  16.92    & 17     &0.093 & 0.483 & 0.517 & 0.513 \\
\hline
\end {tabular}
\end{center}
\vspace*{-0.5cm}
\caption{\small Same as in Table \ref{tab:oxy}
  for the
  $1^-$ states of the various tin isotopes identified by the arrows in
  Fig. \ref{fig:sn}. 
}
\label{tab:sn}
\vspace*{+0.2cm}
\end{table}

\begin{figure}[ht]
\begin{center}
\includegraphics[scale=0.45, angle=0] {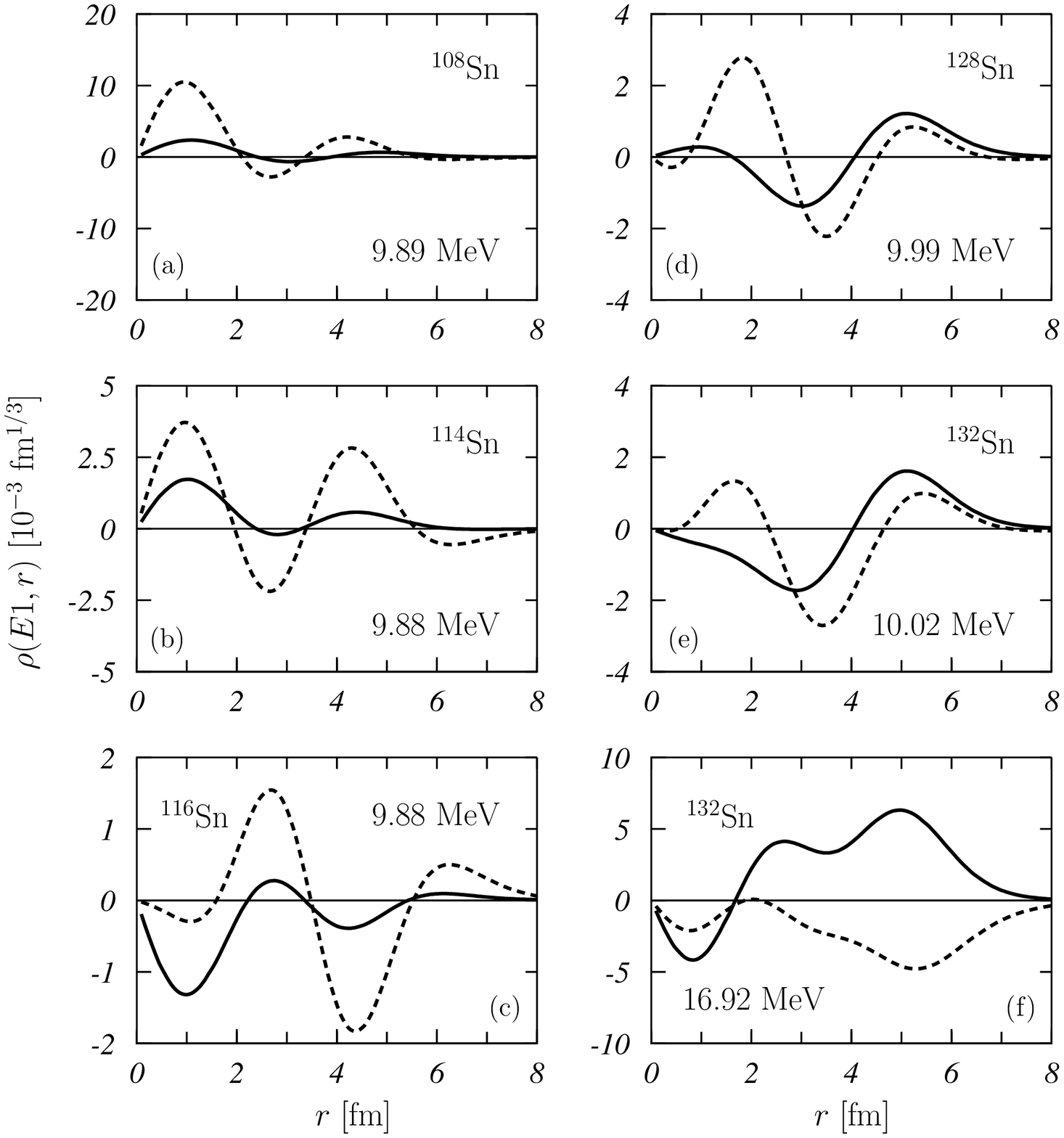} 
\vspace*{-0.3cm}
\caption{\small Same as in Fig.~\ref{fig:toxy}, but
    for the tin isotopes we have studied. 
    }
\label{fig:tsn}
\end{center}
\vspace*{-0.7cm}
\end{figure}

The $B(E1)$ distributions for these isotopes are shown in
Fig. \ref{fig:sn}. For all the nuclei, we found, in addition to the
GDR, also a group of states which appear at lower energies. We give in
Table \ref{tab:sn} the values of the various indexes for some
characteristic states. The states at about 7.8 MeV and 8.4 MeV have
large $B(E1)$ values also in the lighter isotopes. We found it
interesting to follow the development of the state at about 10
MeV. This state has negligible $B(E1)$ values in all the isotopes up
to $^{116}$Sn. Its $B(E1)$ value becomes visible in $^{128}$Sn, about
the 0.2 \% of the global $B(E1)$ contribution, and it further
increases in $^{132}$Sn. With respect to the other tin isotopes, the
additional neutrons of $^{132}$Sn produce a relevant contribution for
this state, as indicated by the relatively large value of \nst shown
in Table \ref{tab:sn}. We show in Fig. \ref{fig:tsn} the transition
densities for these 10 MeV states in the various isotopes, and for the
sake of comparison, also that of the state at the peak of the GDR in
$^{132}$Sn. The relative behavior of the proton and neutron transition
densities clearly shows the $IS$ structure of the states below the GDR
and the $IV$ structure of the 16.92 MeV state in $^{132}$Sn.

\begin{figure}[ht]
\begin{center}
\includegraphics[scale=0.35, angle=0] {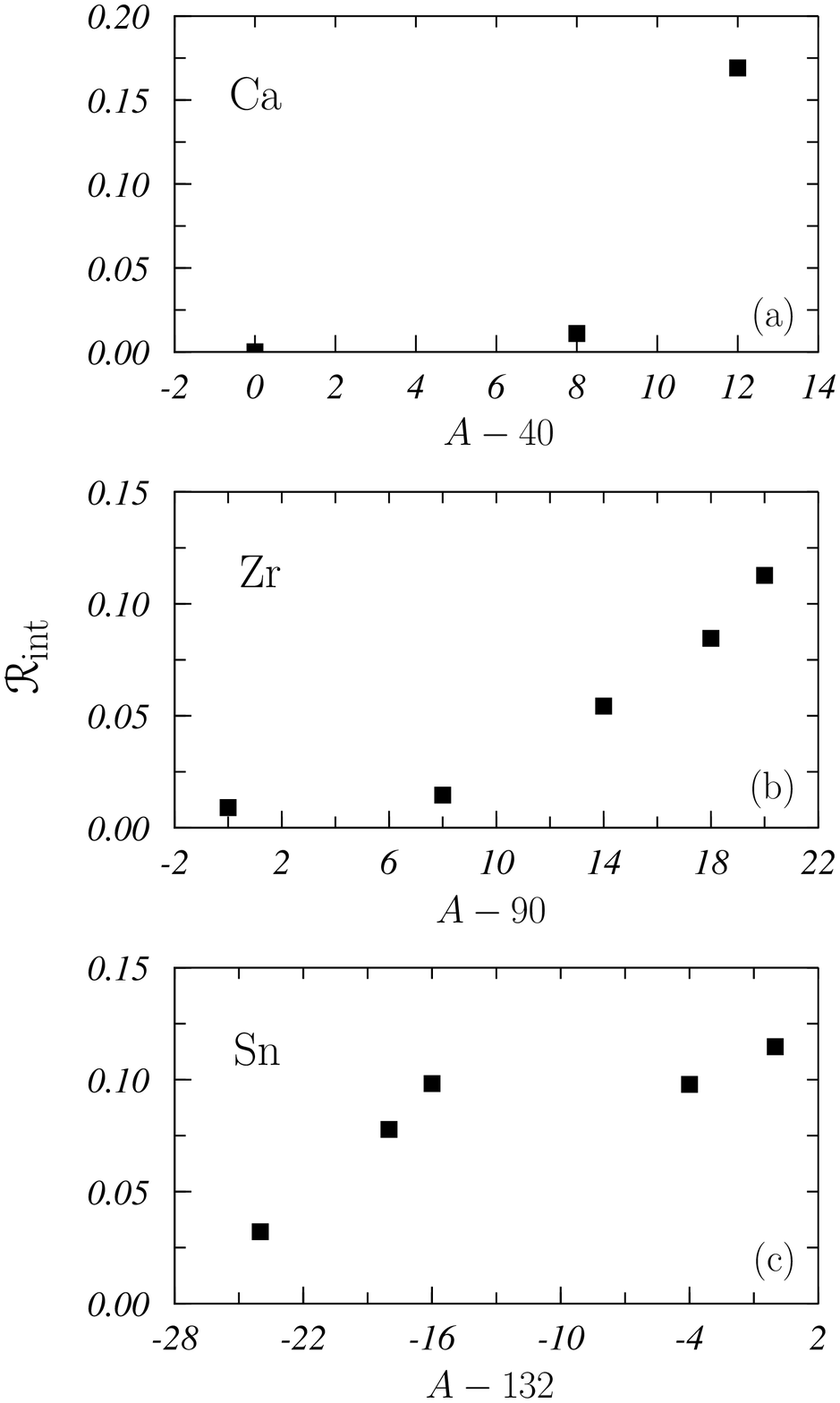} 
\vspace*{-0.3cm}
\caption{\small Ratios $\rbe_{int}$ between the integrated 
    $B(E1)$ values of the PDR, and the
    global $B(E1)$ strength against the number of neutrons in excess with
    respect to the doubly magic core of 
    the Ca, panel (a), Zr, panel (b), 
    and Sn, panel (c), 
    isotopic chains. The $B(E1)$ values of the PDRs have been
    obtained as a sum of all the values below the dashed lines
    indicated in Figs. \ref{fig:ca}, \ref{fig:zr} and \ref{fig:sn}.
    }
\label{fig:int}
\end{center}
\end{figure}

The results presented above indicate that in medium heavy nuclei with
neutron excess, a new type of dipole resonance appears, with the
characteristics we attribute to the PDR, and that the presence of this
resonance becomes more important with the increase of the neutron
number.  We further investigated this point by considering the ratio
$\rbe_{int}$ between the integrated $B(E1)$ low energy strength and the
total strength. We calculated the numerator of $\rbe_{int}$ by summing
all the $B(E1)$ values located below the values indicated by the dashed
lines in Figs. \ref{fig:ca}, \ref{fig:zr} and \ref{fig:sn}, and the
total strength in the denominator by summing over the whole energy
range shown in the figures.

In Fig. \ref{fig:int} $\rbe_{int}$ is plotted against the number of
neutrons in excess with respect to the doubly magic core of the Ca, Zr
and Sn isotopic chains.  These results clearly show an increase of the
relative $E1$ strength in the low energy region with increasing neutron
number.

\begin{figure}[ht]
\begin{center}
\includegraphics[scale=0.35, angle=0] {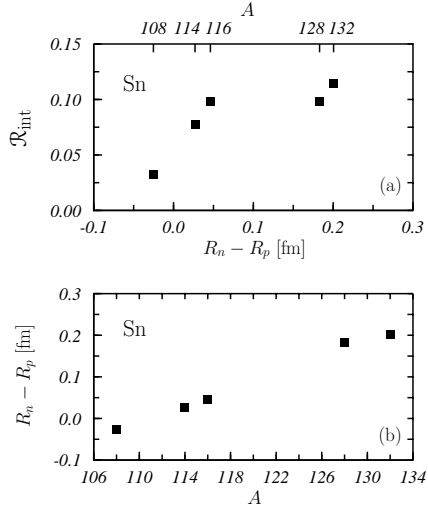} 
\vspace*{-0.3cm}
\caption{\small Panel (a): ratio $\rbe_{int}$ between the integrated 
    $B(E1)$ values of the PDR, and the
    global $B(E1)$ strength against the difference between the proton ($R_p$) 
    and neutron ($R_n$) root mean square radii for, form left to right, 
    A=108, 114, 116, 128 and 132 Sn isotopes. 
    Panel (b): difference $R_n-R_p$ as a function of the mass number A. 
     }
\label{fig:skin}
\end{center}
\end{figure}

The number of Ca isotopes we have considered is too small to allow a
systematic study of the dependence of $\rbe_{int}$ on the neutron
excess, whereas for the Zr chain we observe a monotonic growth.  The
behavior of the Sn isotopic chain is more complicated. A linear growth
is observed for A=108-116, then the value of $\rbe_{int}$ for
$^{128}$Sn is almost equal to that for $^{116}$Sn, and it starts to
grow again, but more slowly, when passing from $^{128}$Sn to
$^{132}$Sn.  This behavior resembles that observed in
Ref.~\cite{pie06} (see also \cite{paa07}), where, within a
relativistic RPA framework, a linear correlation between the ratio of
the low-energy to high-energy dipole strength and the neutron skin of
the Sn isotopes was obtained for $A\leq 120$, followed by an apparent
mild anticorrelation for $120\leq A\leq 132$.  The latter was
attributed to the filling of the $1h_{11/2}$ neutron orbital.

In panel (a) of Fig.~\ref{fig:skin} we show again the ratio
$\rbe_{int}$ for various Sn isotopes, but now as a function of the
neutron skin calculated as a difference between the neutron and
proton root mean square radii, $R_n -R_p$.  The nuclei considered
have, from left to right, $A=108$, $114$, $116$, $128$ and $132$.  In
the lower panel of the figure, we relate these neutron skins to the
isotope mass number.  Although we only consider isotopes with fully
occupied s.p. levels, our results confirm the findings of
Ref.~\cite{pie06}, except for the heaviest isotopes.  In fact, we do
not find any anticorrelation effect since $\rbe_{int}$ mildly
increases in going from $^{128}$Sn to $^{132}$Sn.

\section{Discussion and conclusions}
\label{sec:conc}
We have studied the electric dipole excitation spectra of several
isotopes of oxygen, zirconium and tin nuclei, searching for a possible
appearance of the PDR. The calculations were done for isotopes with
fully occupied s.p. levels using a traditional phenomenological RPA
approach without pairing effects.  In Sect. \ref{sec:model}, we have
critically discussed merits and faults of our approach which we think
to be reliable in predicting position and total strength of the
resonance.

The dipole excited states have been studied by analyzing their
collectivity, their isospin character and the relevance on neutron and
proton p-h excitations. We have defined an index $\doc$, see
Eq. (\ref{eq:doc}), which quantifies the degree of collectivity.  From
the study of the proton and neutron transition densities,
Eq. (\ref{eq:trans1}), we have identified the $IS$ character with the
{\sl in phase} behavior of the two transition densities, while the
{\sl out of phase} behavior indicates the $IV$ character.  We should
mention here that we have also investigated the vorticity of the
excitations \cite{rav87}, as it has been suggested in \cite{rye02},
but we did not find significant differences between the results in the
PDR and GDR regions.  Our work consisted in studying 1$^-$ excitations
in isotopic chains built around doubly closed shell nuclei, to identify
the possible presence of PDR. The signals we searched for identifying
the PDR are high degree of collectivity, $IS$ character and neutron
dominated excitation.

We have first applied our model to oxygen isotopes where we observed
an increase of the $E1$ strength at low energies. These states did not
satisfy our identification criteria.  We observed a fragmentation of
the GDR, rather than the rising of a new type of excitation mode.
This result disagrees with the findings of Ref. \cite{cat97}, where
the calculations were done in a fully self-consistent Hartree-Fock
plus RPA approach with Skyrme-like interactions. The problem is still
open. In any case, we should point out that the oxygen nuclei are
relatively light and in our approach the number of p-h pairs
responsible for the low-lying excitations is so small that it is
difficult to consider these excited states as collective excitation of
the nucleus (see the values of \nst in Table \ref{tab:oxy}).

We found positive results for all the other isotopic chains we have
investigated. Our calculation for the $^{48}$Ca produces strength
around 8.5 MeV in agreement with the experimental findings of Ref.
\cite{har04} and with the results of Ref. \cite{cha94}. Also the
results in tin isotopes show low energy strength and confirm the
experimental finding of Ref. \cite{adr05}. We identify this excitation
as PDR as it has been done in \cite{vre01a}.  This result is in
contrast with the findings of Ref. \cite{sar04}, obtained with
self-consistent calculations which consider also effects beyond the
RPA, in terms of phonon coupling. The authors of Ref. \cite{sar04}
indicate that few p-h excitations are responsible for the low energy
states, while our calculations give quite a relevant collectivity for
these states.

In the study of the zirconium isotopes we have found a handbook
example of the role played by the neutrons in excess. The contribution
of the state at about 8.5 MeV to the $B(E1)$ strength becomes relevant
only in the heavier isotopes, where the neutrons in excess strongly
contribute to the excitation and make it collective.

Our calculations clearly indicate that the appearance of the PDR is a
common feature of medium- heavy-nuclei. By studying the ratio between
the low energy and the total integrated $B(E1)$ strength we have seen
that the relevance of the PDR increases with increasing neutron
number.

\acknowledgments
We thank D. Gambacurta and J. Speth for useful discussions.
This work has been partially supported by the Spanish Ministerio
de Ciencia e Innovaci\'on under contract FPA2008-04688 and by the Junta
de Andaluc\'{\i}a (FQM0220).

\newpage

\end{document}